\documentclass[12pt]{iopart}
\expandafter\let\csname equation*\endcsname\relax
\expandafter\let\csname endequation*\endcsname\relax

\usepackage{iopams} 
\usepackage{amsmath}
\usepackage{bm}
\usepackage{graphicx}
\usepackage{caption}
\usepackage{subcaption}
\usepackage{float} 
\usepackage{mdframed} 
\usepackage{xcolor,tcolorbox}
\usepackage{mathbbol}
\usepackage{pgf,tikz,tikz-cd}
\usetikzlibrary{calc, shapes, intersections,shapes.arrows,arrows.meta,decorations.markings,decorations.pathreplacing,angles,quotes}
\usepackage{cite}
\begin{document}

\title[S Verma et al]{An Inverse Method for the Design of Freeform Double-Reflector Imaging Systems}

\author{Sanjana Verma$^{{1,\star}}$, Lisa Kusch$^{1}$, Martijn J.H. Anthonissen$^{1}$, \newline Jan H.M. ten Thije Boonkkamp$^{1}$, and Wilbert L. IJzerman$^{2,1}$}

\address{$^1$ Department of Mathematics and Computer Science, Eindhoven University of Technology, PO Box 513, 5600 MB Eindhoven, The Netherlands\\
		 $^2$ Signify Research, High Tech Campus 7, 5656 AE Eindhoven, The Netherlands}
\ead{$^{\star}$ s.verma@tue.nl}
\vspace{10pt}

\begin{abstract}
We propose an inverse method to design two-dimensional freeform imaging systems. We present the mathematical model to design a parallel-to-point double-reflector imaging system using inverse methods from nonimaging optics. We impose an imaging condition on the energy distributions at the source and target of the optical system. Our freeform design is compared to the classical Schwarzschild telescope, which is well-known for minimizing third-order aberrations. A raytracer using quasi-interpolation is employed to test the performance of both designs by comparing the spot sizes corresponding to on-axis and off-axis light rays. We show that the inverse freeform design outperforms the classical design.
\end{abstract}

%
%
%
%
%

\section{Introduction}\label{sec: intro}

Freeform imaging systems are increasingly utilized in a broad range of applications enabled by high-performing compact designs \cite{notes, fabian}. Advanced fabrication techniques have bridged the gap between the design and manufacturing of freeform optical systems \cite{notes, janick, fabian}. For designing imaging systems, the goal is to form an ideal image of an object, which means that all light rays from a point in the object plane must converge to exactly one point in the image plane. However, in reality, optical aberrations, i.e., deviations from an ideal image, are inevitable. Therefore, imaging design aims at minimizing aberrations.

Traditional imaging design methods are the so-called \textit{forward} methods like the Simultaneous Multiple Surfaces (SMS) method, achieving favorable designs using aberration theory, and optimization methods for aberration compensation which minimize a merit function dependent on the design parameters of the optical system. \cite{SMS, fabian, korsch, braat}. Recent direct \textit{freeform} imaging design strategies include approaches based on nodal aberration theory and the ``first time right'' design method in which the optical surfaces are calculated by solving differential equations based on Fermat's principle and user-defined conditions like minimal aberrations \cite{NAT, fabian}.

In practice, commercial optical design software containing numerous local and global optimization algorithms is employed for efficient and accurate system design. The optimization algorithms are iterative and inherently pose challenges like the requirement of a good starting guess and the presence of many local minima without a well-defined global minimum\cite{braat,opticaldesignsoftware}. Forward methods like AI-assisted optimization methods in which the solution space of local minima is analyzed to obtain optimal designs and evolutionary algorithms effectively mitigate the challenges mentioned above \cite{evolutionary, AIassistedopt}. 

Alternatively, we aim to develop \textit{inverse} methods for the design of freeform imaging systems. For reflective optical systems, inverse methods in \textit{nonimaging} optics are used to compute the shapes and locations of the freeform reflectors by finding an optical map that converts a given source distribution to a desired target distribution \cite{lotte}.  In \cite{sanjana}, we demonstrated that imaging systems can be designed using inverse nonimaging methods by optimizing the ratio of energy distributions at the source and target. In this paper, we show that the optical map for an aberration-free optical system can be utilized to establish a relation between the energy distributions of an \textit{imaging} system. Using this relation, inverse methods from nonimaging optics can be adapted to design imaging optical systems. 

We will refer to the object and image planes of an imaging system as the source and target planes, respectively, as this terminology is commonly used in inverse methods for nonimaging topics. In nonimaging optics, inverse methods are based on determining an optical map by combining the principles of geometrical optics with the law of conservation of energy. On the other hand, in imaging optics, optical design depends on the optical map that produces minimal aberrations. 
In this paper, we impose the aforementioned optical map and combine it with energy conservation to determine the ratio of energy distributions at the source and target. We then employ inverse methods from nonimaging optics to calculate the shapes of the reflectors. We verify our design with raytracing that relies on a high-order local interpolation method called quasi-interpolation. A local method ensures computational efficiency and a high-order interpolation is required to avoid the dominance of numerical errors in the raytracer.

This paper is organized as follows. In Sect.~\ref{sec: imaging}, the optical Hamiltonian and the condition for the optical map to produce images with zero aberrations are described. In Sect.~\ref{sec: model}, the mathematical model for the design of a two-dimensional (2D) parallel-to-point double-reflector imaging system is presented. Utilizing the classical Schwarzschild telescope, we present the design procedure and the verification method in Sect.~\ref{sec: inverse imaging}-\ref{sec: raytracing}. The performance of the inverse freeform design is compared to the Schwarzschild design in Sect.~\ref{sec: results}, followed by some conclusions and proposed extensions in Sect.~\ref{sec: conclusion}.  

\section{Imaging condition}\label{sec: imaging}


Let the $z$-axis be the optical axis and let the $z$-coordinate be the evolution coordinate as a light ray propagates through an optical system. In the framework of geometrical optics, the so-called phase-space coordinates $q=q(z)$ and $p=p(z)$, completely describe the ray. The position $q$ and momentum $p$ coordinates are the projections of the position vector and the unit direction vector of the ray (multiplied by $n$, which is the refractive index of the medium) on a line $z=\rm{constant},$ respectively. 

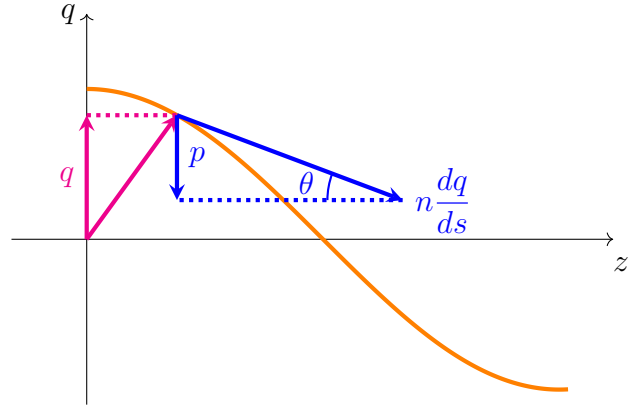
\begin{figure}[H]
    \centering
    \begin{tikzpicture}[scale=2]
    \draw [->] (-0.5,0) -- (3.5,0) node [below] {$\phantom{0}z$};
    \draw [->] (0,-1.1) -- (0,1.5) node [left] {$q$};
    \draw [orange,smooth,samples=100,domain=0:3.2, ultra thick] plot(\x,{cos(((\x))*180/pi)});
	
    \def\z{0.6}; 
	\def\q{{cos(((\z))*180/pi)}};
	\coordinate (P) at (\z,\q);
	
	\draw [magenta, ultra thick, ->,>=stealth] (0,0) -- (P);
	\draw [magenta, ultra thick, ->,>=stealth] (0,0) -- (0,\q) node [midway,left] {$q$};
	\draw [magenta, ultra thick, dotted] (0,\q) -- (P);
	
	\def\dqdz{{-sin(((\z))*180/pi)}};
	\coordinate (Q) at ($(P) + (1.5,\dqdz)$);
	\draw [blue, ultra thick, ->,>=stealth] (P) -- (Q) node [right] {$n \dfrac{dq}{ds}$};
	\draw [blue, ultra thick, ->,>=stealth] (P) -- +(0,\dqdz) node [midway, right] {$p$};
	\draw [blue, ultra thick, dotted] (Q) -- +(-1.5,0);
    \coordinate (R) at ($(Q) + (-1.5,0)$); 
\pic [draw, thick, angle radius=1cm, angle eccentricity=1.3, " $\textcolor{blue}{\theta}$ ", blue] 
    {angle = P--Q--R};

  \end{tikzpicture}
  \caption{An illustration of the phase-space coordinates $(q,p)$ on a line $z=\rm{constant}$, where $s$ denotes the arc-length.}
  \label{fig: q_p 2D}
  \end{figure}

In a 2D optical system, as a ray propagates along the $z$-axis, the rates of change of $q$ and $p$ 
are given by the Hamiltonian system
\begin{subequations}\label{eq: full H system}\begin{align}
&q'=\frac{\partial H}{\partial p},\quad p'=-\frac{\partial H}{\partial q},\label{eq: hamiltonian system}\\
&H(q,p)=-\sigma\sqrt{n^2-p^2},\label{eq: hamiltonian}
\end{align}
\end{subequations}
 where $H$ is the optical Hamiltonian, $p=n\sin\theta$, and $\theta$ is the angle of the tangent vector to the ray with respect to the optical axis (see Fig.~\ref{fig: q_p 2D}). The variable $\sigma$, corresponds to forward $(\sigma=1)$ and backward $(\sigma=-1)$ propagation of the rays. Since, we consider a reflective optical system, $n=1$ in the sequel. 

Let us denote the phase-space coordinates of the incoming and outgoing rays by $\bm{\omega}_{\rm{s}}=(q_{\rm{s}},p_{\rm{s}})^{\rm{T}}$ and $\bm{\omega}_{\rm{t}}=(q_{\rm{t}},p_{\rm{t}})^{\rm{T}}$, where subscripts $\rm{s}$ and $\rm{t}$ denote the source and target, respectively. The phase-space coordinates at the target are connected to the source coordinates by the optical map $\mathcal{M}: \bm{\omega}_{\rm{s}}\mapsto\bm{\omega}_{\rm{t}}$ i.e.,
\begin{equation}\label{eq: optical map M}
\begin{pmatrix}q_{\rm{t}}\\p_{\rm{t}}\end{pmatrix}=\mathcal{M}\begin{pmatrix}q_{\rm{s}}\\p_{\rm{s}}\end{pmatrix}=\begin{pmatrix}M_1(q_{\rm{s}},p_{\rm{s}})\\M_2(q_{\rm{s}},p_{\rm{s}})\end{pmatrix}.
\end{equation}

Let $M$ denote the Jacobian matrix of the optical map $\mathcal{M}$. The optical map $\mathcal{M}$ is governed by the  Hamiltonian system in Eq.~(\ref{eq: full H system}) and therefore $M$ satisfies \cite[p.~164]{lagrangian}
\begin{subequations}
\begin{equation}\label{eq: sympletic}
MJM^{\rm{T}}=J,
\end{equation}
\text{where $J$ is the antisymmetric matrix}\begin{equation} 
J=\begin{bmatrix}0&1\\
-1&0\end{bmatrix},\label{eq: J matrix}\end{equation}\end{subequations}
with properties $J^{-1}=-J$, $J^2=-I,$ $J^{\rm{T}}=-J,$ and $\det(J)=1$. Eq.~(\ref{eq: sympletic}) is the condition for $M$ to be a symplectic matrix and consequently $\det(M)=1$ \cite{symplecticdet1}. This implies that the phase-space optical map $\mathcal{M}$ is symplectic. Symplectic maps are area-preserving \cite[p.~564]{symplecticarea}. This means that as a light beam propagates through the optical system, the area it occupies in phase space, also called \'etendue $\text{d}U=n\,\text{d}q\,\text{d}p=\text{d}q\,\text{d}p$, is conserved. 


 We are interested in designing an optical system with minimum aberrations. So, we identify the optical map $\mathcal{M}$ in this context. An optical map consists of linear and non-linear parts. The underlying mathematical theory is related to Lie transformations and an elaborate description is beyond the scope of this work. A brief overview is as follows and we refer the reader to \cite{aberrationfree} for all relevant details.
 
The optical map $\mathcal{M}$ can be expressed as a concatenation of Lie transformations, under the condition $\mathcal{M}(\bm{0})=\bm{0}.$ The nature of the Lie transformation determines whether it produces a linear or non-linear map when it acts on the source phase-space coordinates. The linear map corresponds to the paraxial optical map, which is free of all aberrations, and the non-linear map includes aberrations. Therefore, the ideal optical map $\mathcal{M}$ would be a linear map. 

We work with the matrix form of the linear map in phase space, i.e.,
\begin{equation}\label{eq: optical map matrix}
    \begin{pmatrix}q_{\rm{t}}\\p_{\rm{t}}\end{pmatrix}
    =\begin{pmatrix}m_{11}&m_{12}\\m_{21}& m_{22}\end{pmatrix}\begin{pmatrix}q_{\rm{s}}\\p_{\rm{s}}\end{pmatrix},
\end{equation}
where $m_{ij}$ for $i,j=1,2$ are constants. Eq.~(\ref{eq: optical map matrix}) is analogous to the conventional matrix formulation of Gaussian optics \cite{hecht}, which describes the linear map in terms of the position and angle of the ray and holds for small angles where $p=\sin\theta\approx\theta.$ 
The reason for choosing the linear map in phase-space coordinates will be apparent in Sect.~\ref{sec: model}.

\section{Mathematical model}
\label{sec: model}

\begin{figure}[htbp]
\centering
\begin{tikzpicture}[scale=0.6]
	\draw  [->] (-1.5,-1.5) node [left] { $z=-l$} -- (7,-1.5) node [below] {$x$};
				\draw [->](-1.5,4.5) node [left] {$z=0$} -- (7,4.5)node [below] {$y$};
				\draw  [->] (5,-2) -- (5,6) node [left] {$z$};	
				\draw [black,line width=2.2pt,name path=reflector1,text width=2.5cm] (-1,0.8) arc(140:110:5) node [left, blue,yshift=-0.7cm,xshift=-1.2cm,align=right] {$\mathcal{R}_1$\\ $z=-l+u(x)$};
				\draw [blue, ultra thick] (-1,-1.5) -- (1.2,-1.5) node [below] {{$\mathcal{S}$}};
				
				\draw [black,line width=2.2pt,name path=reflector2,text width=2.5cm] (2.8,-0.46) arc(270:330:4.1) node [below right, blue] {$\mathcal{R}_2$\\\hspace{-0.7cm} $r=-w\big(\mathbf{\hat{\vphantom{X}t}}\big)\!\cdot{\mathbf{\hat{\vphantom{X}t}}}$};
				\coordinate (T) at (5,4.5);
				
				\def\arrowpos{0.6};
				\path [name path=rayin] (0.5,0) -- (0.5,5);
				\begin{scope}[decoration={markings,mark=at position \arrowpos with {\arrow{stealth}}}]
				\path [name intersections={of=rayin and reflector1, by=A}];
				\draw [orange, ultra thick, postaction={decorate}] (0.5,-1.5) -- (A)  node [midway, left] {${\mathbf{\hat{\vphantom{X}s}}}$};
				\path [name path=rayout] (3.5,-2) -- (T);
				\path [name intersections={of=rayout and reflector2, by=B}];

                \def\arrowpos{0.5};
				\draw [orange, ultra thick, postaction={decorate}] (A) -- (B) node [midway,below left,xshift=0.1cm] {${\mathbf{\hat{\vphantom{X}i}}}$};
				\def\arrowpos{0.55};
				\draw [orange, ultra thick, postaction={decorate}] (B) -- (T) node [midway, left] {${\mathbf{\hat{\vphantom{X}t}}}$};
				\end{scope}	
				\draw (A) node [above,yshift=1mm] {$P_1$};
				\draw (B) node [below] {$P_2$};
				\path [name path=lastpartray] (B) -- (T);
				\path [name path=unitcircle] (T) circle(1cm);
				\path [name intersections={of=lastpartray and unitcircle, by=C}];
				\filldraw [blue] (T) node [above right] {{$\mathcal{T}$}} circle(3pt);
\end{tikzpicture}
\caption{A double-reflector system with a parallel source and a point target.}
\label{fig: 2D optical system}
\end{figure}
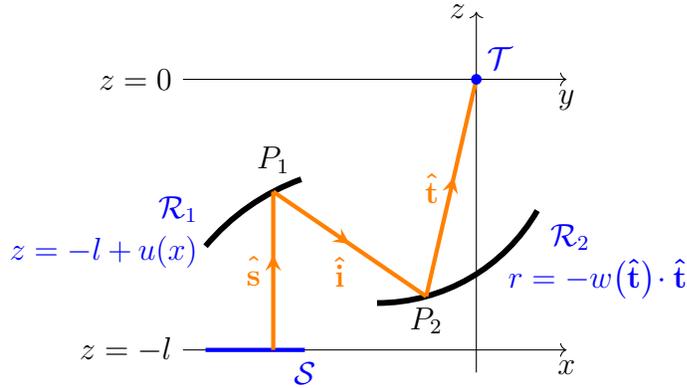

We present the mathematical model for the design of a two-dimensional parallel-to-point freeform imaging system consisting of two reflectors. The model contains four components, outlined as follows. First, the principles of geometrical optics give a \emph{geometrical equation} describing the shapes and locations of the reflectors. The geometrical equation is in terms of the so-called cost function in the framework of optimal transport theory. Second, the law of \emph{conservation of energy} leads to an initial value problem (IVP) for the optical map, which connects the source to the target. This optical map is a function of the ratio of the energy distributions at the source and target. Next, we \emph{impose a linear optical map} in phase space and conclude that a constant energy distribution ratio is required. The geometrical equation has multiple solutions. Finally, we discuss the method for \emph{determining a uniquely defined solution}, which is subsequently used to calculate the shapes of the freeform reflectors.\vspace{1em}\\
\textbf{Cost function formulation:} Consider a 2D optical system consisting of two reflectors $\mathcal{R}_1$, $\mathcal{R}_2$, and the $z$-axis as the optical axis as shown in Fig.~\ref{fig: 2D optical system}. We shall denote unit vectors with a hat (\,$\bm{\hat{}}$\,) throughout this paper. A parallel source in $z=-l$ emits rays with the direction vector $\mathbf{\hat{s}}=(0,1)^{\rm{T}}$ and forms a point target. We choose the point target at the origin $\mathcal{O}$ of the target coordinate system. The rays hit the target plane with the direction vector $\mathbf{\hat{t}}=(t_1,t_2)^{\rm{T}}.$ The source is parametrized by the $x$-coordinate. The unit direction vector $\mathbf{\hat{\vphantom{X}t}}$ is parametrized by the stereographic projection $y$ from the south pole \cite[p.~60]{lotte}. The first reflector, $\mathcal{R}_1$: $z=-l+u(x)$, is defined by the perpendicular distance $u(x)$ from the source and the second reflector, $\mathcal{R}_2$: $\mathbf{r}=-w(\mathbf{\hat{t}})\mathbf{\hat{t}}$, is defined by the radial distance $w(\mathbf{\hat{t}})$ from the target. 

A geometrical description of the aforementioned optical system gives rise to the optimal transport formulation 
 \begin{subequations}
 \begin{equation}
 	u_1(x)+u_2(y)=c(x,y),
 	\label{eq: cost}
 \end{equation} 
where $u_1(x)$ and $u_2(y)$ are related to the locations of the first and second reflectors, respectively, and $c(x,y)$ is the cost function. These functions are given by
\begin{equation}\begin{aligned}
u_1(x)&=\log\left(-\frac{u}{\beta}-\frac{x^2}{2\beta^2}+\frac{V+l}{2\beta}\right),\\
u_2(y)&=\log\left(\frac{\beta}{w}\left(1+y^2\right)-2y^2\right),\\
c(x,y)&=2\log\left(1+\frac{xy}{\beta}\right),
	\label{eq: u1, u2, c}\end{aligned}  
\end{equation}
\end{subequations}
where $V$ and $\beta=V-l>0$ are the optical path length (OPL) and the reduced OPL, respectively. We refer the reader to \cite{sanjana} for details.\vspace{1em}\\
\textbf{Energy conservation:} 
For a finite source or target, the radiance $L$ \cite[p.~17]{carmela} is defined as the radiant flux $\text{d}\Phi$ per unit angle $\text{d}\theta$, and per unit projected length $\cos\theta\,\text{d}\ell$ (perpendicular to the ray)
 \begin{equation}\label{eq: radiant flux defn}
     L(\ell,\theta)=\frac{\text{d}\Phi}{\cos\theta\,\text{d}\ell\,\text{d}\theta},
 \end{equation}
where $\text{d}\ell$ is the infinitesimal segment on the source or target of an optical system. Using Eq.~(\ref{eq: radiant flux defn}), the radiant flux can be written in terms of phase space variables as 
 \begin{equation}\label{eq: radiant flux phase space}
\text{d}\Phi=L(q,p)\,\text{d}q\,\text{d}p=L(q,p)\,\text{d}U
 \end{equation}
 where $\text{d}\ell=\text{d}q$ and $\text{d}p=\cos\theta\,\text{d}\theta$. For the radiance in Eq.~(\ref{eq: radiant flux phase space}), we choose the same notation $L$ as in Eq.~(\ref{eq: radiant flux defn}) for ease of presentation. 

 From Sect.~\ref{sec: imaging}, we recall that in any optical system, the \'etendue $\text{d}U$  is conserved, which implies that a beam of light rays transforms a 2D region in source phase space $(q_{\rm{s}},p_{\rm{s}})$ one-to-one to another 2D region in target phase-space $(q_{\rm{t}},p_{\rm{t}})$ and satisfies $\text{d}U_{\rm{s}}=\text{d}U_{\rm{t}}$, i.e., $\text{d}q_{\rm{s}}\text{d}p_{\rm{s}}=\text{d}q_{\rm{t}}\text{d}p_{\rm{t}}.$ Using Eq.~(\ref{eq: radiant flux phase space}), conservation of energy reads
\begin{equation}\label{eq: energy cons phase space}
\iint_{\mathcal{A}} L_{\mathrm{s}}(q_{\mathrm{s}},p_{\mathrm{s}})\,\mathrm{d}q_{\mathrm{s}}\,\mathrm{d}p_{\mathrm{s}}=	\iint_{\mathcal{M}(\mathcal{A})} L_{\mathrm{t}}(q_{\mathrm{t}},p_{\mathrm{t}})\,\mathrm{d}q_{\mathrm{t}}\,\mathrm{d}p_{\mathrm{t}},
\end{equation}
for any subset $\mathcal{A}\subseteq Q_{\mathrm{s}}\times P_{\mathrm{s}}\subseteq \operatorname{supp}(L_{\rm{s}})$ and image set $\mathcal{M}(\mathcal{A})\subseteq Q_{\mathrm{t}}\times P_{\mathrm{t}}\subseteq \operatorname{supp}(L_{\rm{t}})$, where $\operatorname{supp}(L)$ denotes the support of $L$, $Q$ is the spatial domain, and $P$ is the  momentum domain.

 For a parallel source, $p_{\rm{s}}=0$, and for a point target, $q_{\rm{t}}=0$. Therefore, $L_{\mathrm{s}}(q_{\mathrm{s}},p_{\mathrm{s}})=f(q_{\mathrm{s}})\delta(p_{\mathrm{s}}),$ and $L_{\mathrm{t}}(q_{\mathrm{t}},p_{\mathrm{t}})=g(p_{\mathrm{t}})\delta(q_{\mathrm{t}})$,
where $f(q_{\mathrm{s}})$ is the exitance at the source, $g(p_{\mathrm{t}})$ is the intensity at the target,  $\delta(p_{\mathrm{s}})$ and $\delta(q_{\mathrm{t}})$ are Dirac delta functions. Let $\tilde{m}$ denote the optical map from the spatial source $q_{\rm{s}}$ to the target momentum $p_{\rm{t}},$ where $\tilde{m}(q_{\rm{s}})=M_2(q_{\rm{s}},0)$. We adopt the notation $\tilde{m}$ to avoid confusion. For any subset $\tilde{\mathcal{A}}\subseteq Q_{\mathrm{s}}$ and image set $\tilde{m}(\tilde{\mathcal{A}})\subseteq P_{\mathrm{t}}$, Eq.~(\ref{eq: energy cons phase space}) results in
\begin{equation}\label{eq: energy cons phase space 2}
	 \int_{\tilde{\mathcal{A}}} f(q_{\mathrm{s}})\,\mathrm{d}q_{\mathrm{s}}=	\int_{\tilde{m}(\tilde{\mathcal{A}})}  g(p_{\mathrm{t}})\,\mathrm{d}p_{\mathrm{t}}.
\end{equation}
When $\tilde{\mathcal{A}}=Q_{\mathrm{s}}$ and $\tilde{m}(\tilde{\mathcal{A}})= P_{\mathrm{t}}$, we have global energy conservation, implying that the total flux of the source equals that of the target. Substituting the map $\tilde{m}(q_{\mathrm{s}})=p_{\mathrm{t}}$ in Eq.~(\ref{eq: energy cons phase space 2})  results in
\begin{equation}\label{eq: phase-space ODE}
    \tilde{m}'(q_{\rm{s}})=\pm\frac{f(q_{\rm{s}})}{g(\tilde{m}(q_{\rm{s}}))},
\end{equation}
subject to the transport boundary condition (TBC) $\tilde{m}(\partial Q_{\rm{s}})=\partial P_{\rm{t}}$, which means that the boundary of the source domain is mapped to the boundary of the target domain, implying that all light from the source arrives at the target \cite{TBC}.\vspace{1em}\\ 
\textbf{Optical map for imaging:} As shown above, the optical map depends on the energy distributions at the source and target of the system. We want to find an optical map for imaging optical systems. In Sect.~\ref{sec: imaging}, we discussed the phase-space conditions on the optical map of a perfect-imaging system.
We also saw that the phase-space optical map $\mathcal{M}$ is symplectic and concluded that
\renewcommand{\arraystretch}{1.5}
\begin{equation}\label{eq: jacobian matrix}
  \det  \begin{pmatrix}
  \frac{\partial q_{\rm{t}}}{\partial q_{\rm{s}}}& \frac{\partial q_{\rm{t}}}{\partial p_{\rm{s}}}\\ 
     \frac{\partial p_{\rm{t}}}{\partial q_{\rm{s}}}& \frac{\partial p_{\rm{t}}}{\partial p_{\rm{s}}}
    \end{pmatrix}=1.
\end{equation}
\begin{table}[b]
	\renewcommand{\arraystretch}{1.1}
	\caption{Optical maps with the source and target domains.}
	\normalsize
	\label{table: optical maps}
    \centering
	\begin{tabular}{|l|l|}
		\hline 
       $\mathcal{M}: (q_{\rm{s}},p_{\rm{s}})\mapsto(q_{\rm{t}},q_{\rm{t}})$& Eqs.~(\ref{eq: optical map M}),(\ref{eq: optical map matrix})\\
       $\tilde{m}:q_{\rm{s}}\mapsto p_{\rm{t}}$& Eqs.~(\ref{eq: energy cons phase space 2}),(\ref{eq: phase-space ODE}),(\ref{eq: phase-space linear map})\\
       $m: x\mapsto y$&Eq.~(\ref{eq: y=m(x)})\\
		\hline
	\end{tabular}
\end{table}
For inverse imaging design, we establish an imaging condition on the energy distributions utilizing energy conservation and linearity of the optical map in Eq.~(\ref{eq: optical map matrix}). Given $q_{\rm{t}}=0$ and $p_{\rm{s}}=0$, from Eq.~(\ref{eq: optical map matrix}) we obtain $m_{11}=\partial q_{\rm{t}}/\partial q_{\rm{s}}=0.$ Combining this with Eqs.~(\ref{eq: optical map matrix}) and (\ref{eq: jacobian matrix}) results in
\begin{equation}\label{eq: f/g constant}
\frac{\partial p_{\rm{t}}}{\partial q_{\rm{s}}}=-\frac{1}{m_{12}},
\end{equation}
where $m_{12}$ is a constant. Let $-1/m_{12}=\mathcal{K}$. We know that $\tilde{m}'(q_{\rm{s}})=\partial p_{\rm{t}}/\partial q_{\rm{s}}$. Using Eqs.~(\ref{eq: phase-space ODE}) and (\ref{eq: f/g constant}) result in $\mathcal{K}=\pm f/g.$ Energy distributions $f$ and $g$ are always positive, thus, the ratio $f/g=k>0,$ and $\mathcal{K}=\pm k$. We conclude that the ratio of the energy distributions at the source and target of an imaging system must be a constant, i.e., $f/g=k$. A constant ratio implies that choosing one of the source or target energy distributions as uniform results in the other one also being uniform. 

We now determine the phase-space optical map $p_{\rm{t}}=\tilde{m}(q_{\rm{s}})$ using Eq.~(\ref{eq: phase-space ODE}). As mentioned previously, the TBC holds for Eq.~(\ref{eq: phase-space ODE}). In Sect.~\ref{sec: imaging}, we saw that the phase-space optical map $\mathcal{M}$ for any imaging system satisfies $\mathcal{M}(0)=0$, implying that $\tilde{m}(0)=0$ holds for the phase-space optical map $\tilde{m}$ of our parallel-to-point imaging system. Consequently, when we integrate Eq.~(\ref{eq: phase-space ODE}) to obtain the phase-space optical map $\tilde{m}$, the constant of integration is equal to $0$. The phase-space optical map $p_{\rm{t}}=\tilde{m}(q_{\rm{s}})$ reads
\begin{equation}\label{eq: phase-space linear map}
\tilde{m}(q_{\rm{s}})=\pm kq_{\rm{s}}.\end{equation}

 The optimal transport formulation enables us to design reflectors that connect the source coordinates $x$ to the stereographic target coordinates $y$. Therefore, now we will work with the optical map $y=m(x)$ that maps the source domain $\mathcal{S}$ to the stereographic target domain $\mathcal{T}$. The optical map $y=m(x)$ can be determined from the phase-space optical map $p_{\rm{t}}=\tilde{m}(q_{\rm{s}})$ in Eq.~(\ref{eq: phase-space linear map}). The stereographic projection $y$ from the south pole can be written in terms of $p_{\mathrm{t}}$ by the relation \cite{sanjana}
 \begin{equation}\label{eq: y in terms of p}
     y=\frac{p_{\mathrm{t}}}{1+\sqrt{1-p_{\mathrm{t}}^2}}.
 \end{equation}
We substitute $q_{\rm{s}}=x$ in Eq.~(\ref{eq: phase-space linear map}). Subsequently, using Eqs.~(\ref{eq: phase-space linear map})-(\ref{eq: y in terms of p}), we determine the optical map $y=m(x)$ 
\begin{equation}\label{eq: y=m(x)}
    m(x)=\pm\frac{kx}{1+\sqrt{1-(kx)^2}}.
\end{equation}
We provide an overview of the optical maps used in various equations and their corresponding domains in Table~\ref{table: optical maps}.\vspace{1em}\\
\textbf{Freeform reflectors:} 
\newcommand{\myunderset}[2]{\underset{#1}{#2}}
Eq.~(\ref{eq: cost}) has many possible solutions, but we restrict ourselves to the following solution pairs \cite[Ch.~4]{yadav} 
  	\begin{subequations}\label{eq: concave/convex}
    \begin{alignat}{2}
  &\text{c-convex:}	\quad	u_1(x)&=\myunderset{y\in\mathcal{T}}\max\big(c(x,y)-u_2(y)\big), \quad	u_2(y)&=\myunderset{x\in\mathcal{S}}\max\big(c(x,y)-u_1(x)\big),
  	   \label{eq: convexsoln}\\
         &\text{c-concave:}	\quad	u_1(x)&=\myunderset{y\in\mathcal{T}}\min\big(c(x,y)-u_2(y)\big), \quad	u_2(y)&=\myunderset{x\in\mathcal{S}}\min\big(c(x,y)-u_1(x)\big).
  	   \label{eq: concavesoln}
       \end{alignat}
  	\end{subequations}
  	Eq.~(\ref{eq: concave/convex}) requires $x$ to be a stationary point of $c(\cdot,y)-u_1$. As a result, a necessary condition for the existence of the solution is
     \begin{equation}\label{eq: necessary condition}
     c_x(x,y)-u_1'(x)=0.
      \end{equation}
Substituting the map $y=m(x)$ in Eq.~(\ref{eq: necessary condition}) and once more differentiating with respect to $x$ leads to
   \begin{equation}\label{eq: double derivative}
    c_{xy}(x,m(x))m'(x)=u_1''(x)-c_{xx}(x,m(x)).
   \end{equation}

Eq.~(\ref{eq: double derivative}) provides the criterion for determining the nature of the solution. A sufficient condition for a c-convex or c-concave solution is that $c_{xx}(x,m(x))-u_1''(x)$ is negative or positive, respectively. From Eq.~(\ref{eq: u1, u2, c}), we can show that $c_{xy}(x,m(x))>0$, implying that the type of solution obtained in our case depends on the sign of $m'(x).$ This means that we obtain a c-convex solution if $m'(x)>0$ and a c-concave solution if $m'(x)<0.$

Next, we substitute $y=m(x)$ in Eq.~(\ref{eq: necessary condition}) and fix the distance $u_0$ of the center of the first reflector to the source. This leads to an IVP enabling us to calculate $u_1$. We then calculate $u_2$ from Eq.~(\ref{eq: cost}). Subsequently, Eq.~(\ref{eq: u1, u2, c}) results in
   \begin{subequations}\label{eq: reflector shapes}
   \begin{align}
       	u(x)&= -\beta\exp(u_1(x))-\frac{x^2}{2\beta}+\frac{V+l}{2},\label{eq: reflector1 u(x)}\\
	    w(\bm{\hat{t}})&=\beta\left(1+m(x)^2\right)\left(\exp(u_2(m(x)))+2m(x)^2\right)^{-1}.
    \label{eq: reflector2 w(y)}
    \end{align}
   \end{subequations}
A complete derivation is given in \cite{sanjana}. 

\section{Formulation of inverse freeform imaging design}\label{sec: inverse imaging}

In Sect.~\ref{sec: model}, we presented the mathematical model to design freeform imaging systems. Now, we formulate specific choices for various input parameters in the inverse model. These choices depend on a classical imaging design, which will be explained later in this section.

 From Eq.~(\ref{eq: phase-space linear map}), it follows that computing the phase-space map $\tilde{m}\left(q_{\rm{s}}\right)=p_{\rm{t}}$ depends on the value of the ratio $k$ of the energy distributions and the positive or negative sign for the map. The value of the constant $k$ can be determined using energy conservation given by Eq.~(\ref{eq: energy cons phase space 2}). This requires specifying the TBC $\tilde{m}(\partial Q_{\rm{s}})=\partial P_{\rm{t}}$. 
The choice of TBC enables us to design imaging systems using the inverse nonimaging strategy by limiting the spread of light rays, thereby directing all energy to a specified stereographic target domain. We will discuss each of the above sequentially. 

In nonimaging optics, the source and target domains can be chosen freely, implying that energy is conserved on arbitrary domains. However, in imaging systems, this choice is restricted. Classical imaging designs minimize aberrations and provide an estimate of the target domain where all energy is concentrated for a specified source. We know that these domains are characterized by minimal aberrations. Consequently, they are an optimal choice for imaging design.

We aim to design symmetric reflectors as even-order transverse ray aberrations are eliminated in such systems \cite{evenorderaberrations}. The ratio of energy distributions is an even function since it is a constant. Consequently, the optical map is symmetric, leading to symmetric reflectors. Hence, we work with symmetric source and stereographic target domains. An extensive proof is presented in \cite{sanjana}.

For classical imaging design, we choose the Schwarzschild telescope, which consists of aspheric reflectors and produces minimal third-order aberrations \cite{korsch}. Let $\alpha$ denote the angle (measured counterclockwise) that a set of parallel rays from the source makes with the optical axis. If we raytrace the classical design for $\alpha=0^\circ,$ for a symmetric source domain $Q_{\rm{s}}=\left[-a,a\right]$, we obtain the target domain $P_{\rm{t}}=\left[-b,b\right]$. 
These source and target domains are employed to enforce the TBC, which is used to calculate $k$ from Eq.~(\ref{eq: energy cons phase space 2}).
\begin{table}[t]
	\renewcommand{\arraystretch}{1.3}
	\caption{A summary of the parameters and specific choices in the design method.}
	\normalsize
	\label{table: imaging design}
    \centering
	\begin{tabular}{|l|l|}
		\hline 
        Coordinate system & Phase-space coordinates: $q_{\rm{s}}$, $p_{\rm{t}}$\\
        Inputs for inverse methods  &  A linear phase-space optical map $\tilde{m}:Q_{\rm{s}}\rightarrow P_{\rm{t}}$\\
        \, &  Energy distributions: $f(q_{\rm{s}})$, $g(p_{\rm{s}})$\\
        Relation for imaging design & $f/g=k$ \\
        Transport boundary condition & Obtained from classical design\\
       System layout & Obtained from classical design\\
       Sign of the optical map & Obtained from classical design\\
       Optical map for output computations & $m:\mathcal{S}\rightarrow \mathcal{T}$\\
       Output & Freeform imaging reflectors\\
		\hline
	\end{tabular}
\end{table}

Next, we choose the sign for the phase-space optical map $p_t=\tilde{m}(q_{\rm{s}})$ in Eq.~(\ref{eq: phase-space linear map}). From raytracing, we observe that for the Schwarzschild telescope, the left boundary of the source domain is mapped to the right boundary of the target domain and vice versa. Since the optical map $\tilde{m}$ is symmetric about the optical axis, this implies that all rays generated from the source with positive $q_{\rm{s}}-$values reach the target with negative $p_{\rm{t}}-$values. From Eq.~(\ref{eq: phase-space linear map}), we conclude that the negative sign holds. 

We discuss the sign of the optical map $y=m(x)$, which is finally used to calculate the shapes of the reflectors. From Eqs.~(\ref{eq: phase-space linear map})-(\ref{eq: y=m(x)}), we observe that the optical map $m$ depends on the sign of the phase-space optical map $\tilde{m}$. Therefore, due to our choice of the negative sign in Eq.~(\ref{eq: phase-space linear map}), the negative sign also holds for the optical map $m$.

As previously described in Sect.~\ref{sec: model}, the sign of $m'(x)$ helps us determine whether our solution is c-concave or c-convex. We differentiate Eq.~(\ref{eq: y=m(x)}) with respect to $x$ and observe that the sign of $m'(x)$ is same as the sign of the optical map $y=m(x).$ It follows that $m'(x)<0$ and we conclude that our optical system has a c-concave solution. In Table~\ref{table: imaging design}, we summarize the specific choices for various parameters used in the mathematical model for the design of freeform imaging systems.

\section{Quasi-interpolation for raytracing}\label{sec: raytracing}
\begin{figure}[htbp]
\centering
\hspace{9em}
	\begin{tikzpicture}
		\colorlet{rectangleboundary}{blue}
		\colorlet{rectanglefillcolor}{blue!10}
		
		\node[draw=rectangleboundary, thick, fill=rectanglefillcolor, ellipse, minimum width=3cm, minimum height=0.5cm] (ovalstart) at (0,0.7) {\small \textcolor{black}{Start}};
		
		\coordinate (A1) at (-2.3,-0.3);
		\coordinate (B1) at (-2.5,-1.4);
		\coordinate (C1) at (2.4, -1.4);
		\coordinate (D1) at (2.6, -0.3);
		
		\draw [rectangleboundary,fill= rectanglefillcolor, thick] (A1) -- (B1) -- (C1) -- (D1) -- cycle;

		\node[align=center,text centered, text width=5.5cm] (diamond) at (0.1, -0.9) {\small\textcolor{black}{ Read data points for $\mathcal{R}_1,$$\mathcal{R}_2$} \\ \vspace{-0.2em} {Generate rays from $\mathcal{S}$}};
		   
        \draw[-{Triangle[width=6pt,length=4pt]}, line width=2pt] (0,0.3) -- (0,-0.3); 
        
		\node[draw=rectangleboundary, thick, fill=rectanglefillcolor, rounded corners=2pt, minimum width=4.5cm, minimum height=0.5cm] (rect2) at (0,-2.4) {\small \textcolor{black}{Approximate $\mathcal{R}_1$, $\mathcal{R}_2$}};
		\draw[-{Triangle[width=6pt,length=4pt]}, line width=2pt] (0,-1.4) --(rect2.north);
		
		\node[draw, rectangleboundary,thick, fill=rectanglefillcolor, rounded corners=2pt, minimum width=4.5cm, minimum height=0.5cm] (rect3) at (0,-3.7) {\small \textcolor{black}{Calculate $P_1$, Approximate $\mathbf{\hat{n}}$ at $P_1$}};
		
        \draw[decorate, thick,decoration={brace, mirror, amplitude=6pt}] (3.2,-4.1) -- (3.2,-2.1) node[right,align=center,text={orange}] at (3.4,-3.1) {{\small \bf Quasi-interpolation}};
		\draw[-{Triangle[width=6pt,length=4pt]}, line width=2pt] (rect2.south) -- (rect3.north);
		
		\node[draw=rectangleboundary,thick, fill=rectanglefillcolor, rounded corners=2pt, minimum width=4.5cm, minimum height=0.5cm] (rect4) at (0,-5) {\small \textcolor{black}{ Calculate $\mathbf{\hat{i}}=\mathbf{\hat{s}}-2(\mathbf{\hat{s}}\cdot\mathbf{\hat{n}})\mathbf{\hat{n}}$}};
				
		\draw[-{Triangle[width=6pt,length=4pt]}, line width=2pt] (rect3.south) -- (rect4.north);

        \node[draw, rectangleboundary,thick, fill=rectanglefillcolor, rounded corners=2pt, minimum width=4.5cm, minimum height=0.5cm] (rect5) at (0,-6.3) {\small \textcolor{black}{Calculate $P_2$, Approximate ${\bm{\mathit{\hat{n}}}}$ at $P_2$}};

        \draw[decorate, thick,decoration={brace, mirror, amplitude=3pt}] (3.2,-6.7) -- (3.2,-5.9) node[right,align=center,text={orange}] at (3.4,-6.3) {{\small \bf Quasi-interpolation}};
        
		\draw[-{Triangle[width=6pt,length=4pt]}, line width=2pt] (rect4.south) -- (rect5.north);
		\node[draw, rectangleboundary,thick, fill=rectanglefillcolor, rounded corners=2pt, minimum width=4.5cm, minimum height=0.5cm] (rect6) at (0,-7.6) {\small \textcolor{black}{ Calculate $\mathbf{\hat{t}}=\mathbf{\hat{i}}-2(\mathbf{\hat{i}}\cdot\mathbf{\hat{n}})\mathbf{\hat{n}}$}};
		\draw[-{Triangle[width=6pt,length=4pt]}, line width=2pt] (rect5.south) -- (rect6.north);
		
		\node[draw, rectangleboundary,thick, fill=rectanglefillcolor, rounded corners=2pt, minimum width=4.5cm, minimum height=0.5cm] (rect7) at (0,-8.9) {\small \textcolor{black}{ Calculate $Y_{\rm{t}}$}};
		\draw[-{Triangle[width=6pt,length=4pt]}, line width=2pt] (rect6.south) -- (rect7.north);

		\node[draw=rectangleboundary, thick, fill=rectanglefillcolor, ellipse, minimum width=3cm, minimum height=0.5cm] (ovalend) at (0,-10.2) {\small \textcolor{black}{End}};
		\draw[-{Triangle[width=6pt,length=4pt]}, line width=2pt] (rect7.south) -- (ovalend.north);
	
	\end{tikzpicture}
    \caption{A schematic of a raytacer.}
    \label{fig: raytracer}
\end{figure}
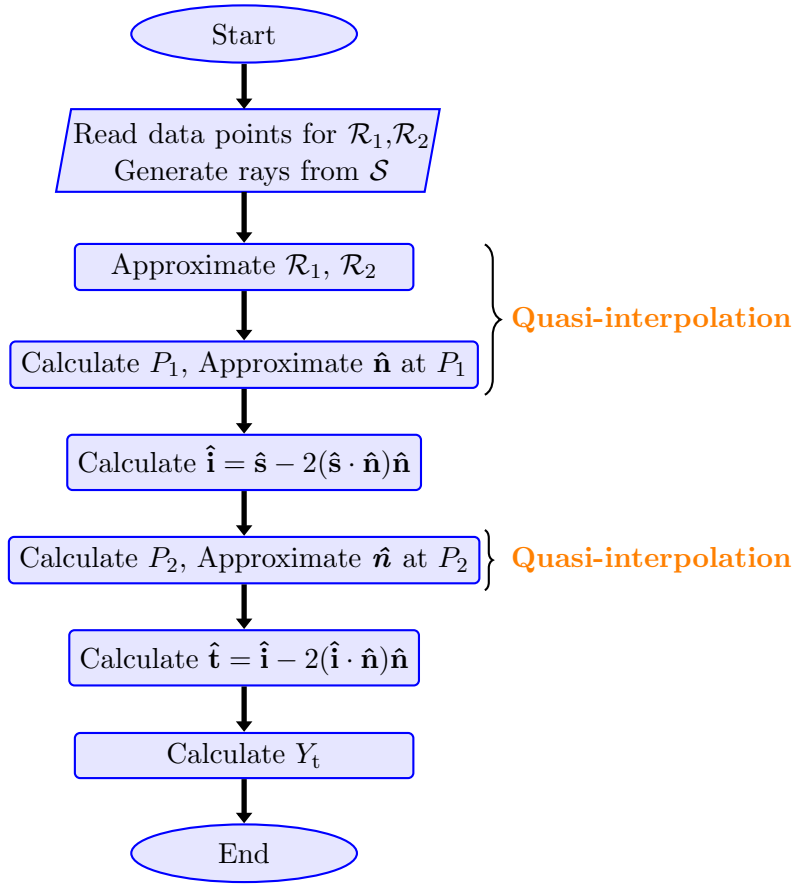

We verify our freeform design with a raytracer by calculating the root-mean-square (RMS) spot size $W$ corresponding to an incoming parallel beam of light
\begin{equation}\label{eq: RMS spot size}
W=\sqrt{\rm{Var}[Y_{\rm{t}}]},
\end{equation}
where $Y_{\rm{t}}$ denotes the landing position of a ray at the target screen. For any ray, a raytracer determines the path of the ray through the optical system by finding the points of intersection on the reflectors and applying the law of reflection to find the ray directions. The vectorial law of reflection \cite[p.~22-24]{lotte} is given by $\mathbf{\hat{t}}=\mathbf{\hat{s}}-2(\mathbf{\hat{s}}\cdot\mathbf{\hat{n}})\mathbf{\hat{n}}$, where $\mathbf{\hat{s}}$ and $\mathbf{\hat{t}}$ denote the incident and reflected unit direction vectors, respectively, and $\mathbf{\hat{n}}$ denotes the unit normal vector of the reflector at the point where the ray intersects it.

To estimate the spot size accurately, we have to determine an accurate target coordinate $Y_{\rm{t}}$ for each ray. The accuracy of $Y_{\rm{t}}$ depends on the accuracy of the normals on both reflectors (refer to Fig.~\ref{fig: raytracer}). The reflectors are calculated as point clouds, which are then used to approximate the normals. Small inaccuracies in the calculation of the normals can result in highly inaccurate spot sizes, especially if we want to precisely estimate the spot size using a small number of rays. Additionally, we want to approximate the normals with a low computational cost. Therefore, our goal is to approximate the derivatives of the reflectors smoothly utilizing a coarse reflector grid. 

We present an approximation method for a real-valued even function $h$ and its first derivative. The same method is then applied for approximating both freeform reflectors, $\mathcal{R}_1: z=-l+u(x)$, and $\mathcal{R}_2: \mathbf{r}=-w(\mathbf{\hat{t}})\mathbf{\hat{t}}$, and their corresponding normals. Our design consists of reflectors symmetric about the central ray, which means the reflectors are symmetric about $x=0$. Therefore, we also tailor our approximation scheme to be symmetric.

First, we introduce the quasi-interpolation (QI) method \cite{QI}. QI is an easy-to-construct local method that smoothly approximates a real-valued function and its derivatives without solving a large system of equations. Our aim is to construct a symmetric QI scheme for a real-valued function $h\in\mathcal{P}_d$, where $\mathcal{P}_d$ is the space of polynomials of total degree at most $d$. Let $B_{i,d,\bm{\xi}}$ denote the $i^{th}$ B-spline of degree $d$ defined on an open knot sequence $\bm{\xi}$ \cite{QIlychee}
\begin{equation}
\bm{\xi}:=\{\xi_j\}_{j=1}^{n+d+1}=\{\xi_{1}=\cdots=\xi_{d+1}<\xi_{d+2}\leq\cdots\leq\xi_{n}<\xi_{n+1}=\cdots=\xi_{n+d+1}\},
\end{equation}
where $n$ is the total number of B-splines defined on $\bm{\xi}$. 
For $i=1,2,\ldots,n$, $B_{i,d,\bm{\xi}}$ is identically zero if $\xi_{i+d+1}=\xi_i,$ and is otherwise defined recursively \cite{QIlychee} by 

\begin{subequations}\label{eq: b-spline}
\begin{align}
   B_{i,d,\bm{\xi}}(x)&=\frac{x-\xi_i}{\xi_{i+d}-\xi_i}B_{i,d-1,\bm{\xi}}(x)+\frac{\xi_{i+d+1}-x}{\xi_{i+d+1}-\xi_{i+1}}B_{i+1,d-1,\bm{\xi}}(x),\\
\intertext{with}  B_{i,0,\bm{\xi}}(x)  &=\begin{cases}1 & \text{if } x\in[\xi_i,\xi_{i+1}),\\
0 & \text{otherwise}.
\end{cases}
    \end{align}
\end{subequations}
A B-spline quasi-interpolant \cite[Chap.~2]{QI} is an operator of the form
\begin{equation}\label{eq: QI defn}
    	Q_d\,[h]=\sum_{i=1}^{n}\lambda_{i,d,\bm{\xi}}(h)B_{i,d,\bm{\xi}},
\end{equation}
where 
$\lambda_{i,d,\bm{\xi}}(h)$ are coefficients defined as a linear combination of the values of $h$ at some points in the neighborhood of the support of $B_{i,d,\bm{\xi}}$. This implies that the approximation of $h$ at a point depends on the values of $h$ in its neighborhood only. Therefore, the approximation requires less data points compared to standard interpolation techniques, implying that a coarse grid can be used to approximate the reflectors.

The normals of the reflectors are calculated using the derivative $\text{D}h$ given by \cite{QIlychee}  
\begin{equation}\label{eq: QI derivative}
\text{D}h=\sum_{i=2}^{n} c_iB_{i,d-1,\bm{\xi}}, \quad c_i=d\frac{\lambda_i-\lambda_{i-1}}{\xi_{i+d}-\xi_i}.
\end{equation}
In our computations, we assume that the values of $h$ are known at equidistant grid points. The approximation order is $\mathcal{O}$$((\Delta\xi)^{d+1})$, where $\Delta\xi$ is the distance between the uniformly spaced grid points at which the function values are known. 

For a good approximation, we require that the QI scheme must be exact on the space of polynomials of at most degree $d$, i.e., $Q_d[s]=s$, for all $s\in\mathcal{P}_d$ \cite[Chap.~2]{QI}. Polynomials can be expressed as a linear combination of B-splines of degree $d$ using the Marsden identity \cite[Chap.~4]{QI}. Based on the above, the following QI approximation \cite{QIlychee} holds for $h\in\mathcal{P}_d$
\begin{subequations}\label{eq: b-spline coeff main}
			\begin{align}
				h(x)&\approx\sum_{i=1}^{n}\lambda_{i,d,\bm{\xi}}(h)B_{i,d,\bm{\xi}}(x), \quad x\in[\xi_{d+1},\xi_{n+1}],\label{eq: rep of b-spline coeff}\\ 
			\lambda_{i,d,\bm{\xi}}(h)&=\begin{cases}\frac{1}{d!}\sum_{k=\mu_i}^{d}(-1)^{d-k}\text{D}^{k}\Psi_{i,d,\bm{\xi}}(\tau_i)\Delta_{+}^{d-k}h(\tau_i), \quad \tau_i=\xi_i,\\
					\frac{1}{d!}\sum_{k=\mu_i}^{d}(-1)^{d-k}\text{D}^{k}\Psi_{i,d,\bm{\xi}}(\tau_i)\Delta_{}^{d-k}h(\tau_i), \quad \xi_i<\tau_i<\xi_{i+d+1},\\
					\frac{1}{d!}\sum_{k=\mu_i}^{d}(-1)^{d-k}\text{D}^{k}\Psi_{i,d,\bm{\xi}}(\tau_i)\Delta_{-}^{d-k}h(\tau_i), \quad \tau_i=\xi_{i+d+1},
				\end{cases}  \label{eq: rep of b-spline coeff 1}\\ 
                  \Psi_{i,d,\bm{\xi}}(\tau_i)&=\prod_{k=1}^{d}(\tau_i-\xi_{i+k}),
			\end{align}
		\end{subequations}
	where  $\text{D}^k$ denotes the $k^{\mathrm{th}}$ derivative, $\Psi_{i,d,\bm{\xi}}$ is the so-called dual polynomial, $\tau_i\in(\xi_i,\xi_{i+d+1})$, and $\mu_i\geq0$ is the number of times $\tau_i$ appears in $\xi_{i+1},\ldots,\xi_{i+d}$. In Eq.~(\ref{eq: rep of b-spline coeff 1}), $\Delta^{d-k}h(\tau_i)$ are calculated using central finite differences. Also, $\Delta_{+}^{d-k}h(\tau_i)$ and $\Delta_{-}^{d-k}h(\tau_i)$ in Eq.~(\ref{eq: rep of b-spline coeff 1}) are calculated using forward and backward asymmetric finite difference schemes, respectively.
    
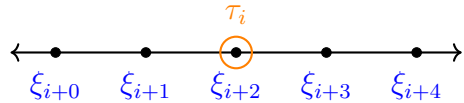
\begin{figure}[htbp]
 \centering
 \begin{tikzpicture}[scale=0.6]

\draw[thick,<->] (0,0) -- (10,0);
	
\foreach \i in {0,1,2,3,4} {
	\node[below, yshift=-0.1cm] at (\i*2+1,0) {\textcolor{blue}{$\xi_{i+\i}$}}; 
 }
	
\foreach \i in {0,1,2,3,4} {
	\filldraw (\i*2+1,0) circle (3pt); 
 }
	
\draw[thick,orange] (5,0) circle (0.35cm);
	
\node[above] at (5,0.4) {\textcolor{orange}{$\tau_{i}$}};
\end{tikzpicture}
\caption{ Choice of $\tau_{i}$ for a symmetric approximation.}\label{fig: tau choice}
\end{figure}
We discuss the implementation of the aforementioned formulas in Eq.~(\ref{eq: rep of b-spline coeff 1}). Without loss of generality, we assume that the reflector grid is known at $\bm{\xi}$, which means that the function values are known at the knot vector, i.e., $h(\xi_i)=h_i$, for $1\leq i\leq n+d+1$. Since the reflectors are symmetric about $x=0$, we find that the values $h_i$ are symmetric about $x=0$ and that the  knot vector $\bm{\xi}$ satisfies $\xi_j=-\xi_{n+d-j+2}$, for $1\leq j\leq \lfloor \frac{n+d+1}{2} \rfloor$, where $\lfloor\cdot\rfloor$ denotes the floor function. From Eqs.~(\ref{eq: QI defn}) and (\ref{eq: rep of b-spline coeff}), we observe that $Q_d[h(x)]=Q_d[h(-x)]$ holds under the following conditions:
\begin{enumerate}
    \item $B_{j,d,\bm{\xi}}(-x)$=$B_{n-j+1,d,\bm{\xi}}(x)$, for all $x\in[\xi_{d+1},\xi_{n+1}]$, and $1\leq j\leq\lfloor n/2\rfloor$,
    
    \item $\lambda_{i,d,\bm{\xi}}$ are symmetric throughout the domain of the knot vector $\bm{\xi}$.
\end{enumerate}
We choose B-splines $B_{i,3,\bm{\xi}}$ of degree $3$ 
for our computations. Each B-spline is a piecewise cubic polynomial which is non-zero over four subintervals of the knot vector. Uniformly spaced knots ensure that the recursive formula given by Eq.~(\ref{eq: b-spline}) to calculate each B-spline can be applied uniformly on the entire domain. Since the knots are symmetric, the subintervals on which the B-splines are defined are also symmetric. This ensures the symmetry of B-splines throughout the domain.

Since function values $h_i$ are symmetric, symmetry for the coefficients $\lambda_{i,d,\bm{\xi}}$ depends on our choice of $\tau_i$. We now prove that our specific choice of $\tau_i$ ensures a symmetric approximation. $B_{i,3,\bm{\xi}}$ is non-zero in the interval $\left[\xi_i,\xi_{i+4}\right]$ for $1\leq i\leq n$. Each B-spline depends on $5$ knots. As shown in Fig.~{\ref{fig: tau choice}}, we choose $\tau_i=\xi_{i+2}$, which is the center knot of the interval $[\xi_i,\xi_{i+4}]$, for $1\leq i\leq n$. Since $\bm{\xi}$ is symmetric, for $j=1,\ldots,\lfloor n/2\rfloor$, the following holds:
 \begin{subequations}\label{eq: sign psi}\begin{align}
        \tau_j&=-\tau_{n-j+1},\\
 		\Psi_{i,d,\bm{\xi}}\left(\tau_j\right)&=\Psi_{i,d,\bm{\xi}}(-\tau_{n-j+1})=0,\\
 		\text{D}\Psi_{i,d,\bm{\xi}}\left(\tau_j\right)&=\text{D}\Psi_{i,d,\bm{\xi}}(-\tau_{n-j+1}),\\
 		\text{D}^2\Psi_{i,d,\bm{\xi}}\left(\tau_j\right)&=-\text{D}^2\Psi_{i,d,\bm{\xi}}(-\tau_{n-j+1}),\\
 		\text{D}^3\Psi_{i,d,\bm{\xi}}\left(\tau_j\right)&=\text{D}^3\Psi_{i,d,\bm{\xi}}(-\tau_{n-j+1}).
 	\end{align}
 \end{subequations}
Also, for $i=1,\ldots,n$, we know that
 \begin{subequations}\label{eq: sign f}
 \begin{align}
 	h\left(-\tau_i\right)&=h\left(\tau_i\right),\\
\Delta h\left(-\tau_i\right)&=-\Delta h\left(\tau_i\right),\\
\Delta^2h\left(-\tau_i\right)&=\Delta^2 h\left(\tau_i\right),\\
\Delta^3h\left(-\tau_i\right)&=-\Delta^3h\left(\tau_i\right).
	\end{align}
\end{subequations}
Substituting Eqs.~(\ref{eq: sign psi})-(\ref{eq: sign f}) in Eq.~(\ref{eq: rep of b-spline coeff 1}), we conclude that the  coefficients $\lambda_{i,d,\bm{\xi}}\left(h\right)$ are symmetric about $x=0.$ Therefore, the approximation $h$ is even and symmetric about $x=0.$ Utilizing the approximation scheme described above yields a smooth and symmetric approximation of the reflectors and their normals.
 
\section{Numerical results}\label{sec: results}

We compare the RMS spot sizes for beams of on-axis and off-axis parallel rays generated from a source of unit length that pass through a Schwarzschild telescope and an inverse freeform telescopic system. To ensure a fair comparison, we choose the same layout for both optical systems, which means that we fix the distance $D_{\rm{v}}=10$ between the vertices of both reflectors and the location of the source $z=-30$. In inverse freeform design, the distance $u_0$ between the vertex of the first reflector and the source affects the layout. We will choose it equal to the location of the vertex of the first reflector of the Schwarzschild telescope. We summarize the procedure for obtaining the numerical results along with the chosen design parameters.
    
\begin{figure}[h]
		\centering
    \begin{minipage}[b]{0.6\textwidth}
	    \includegraphics[trim={0cm 0cm 0cm 0cm},clip, width=\textwidth]{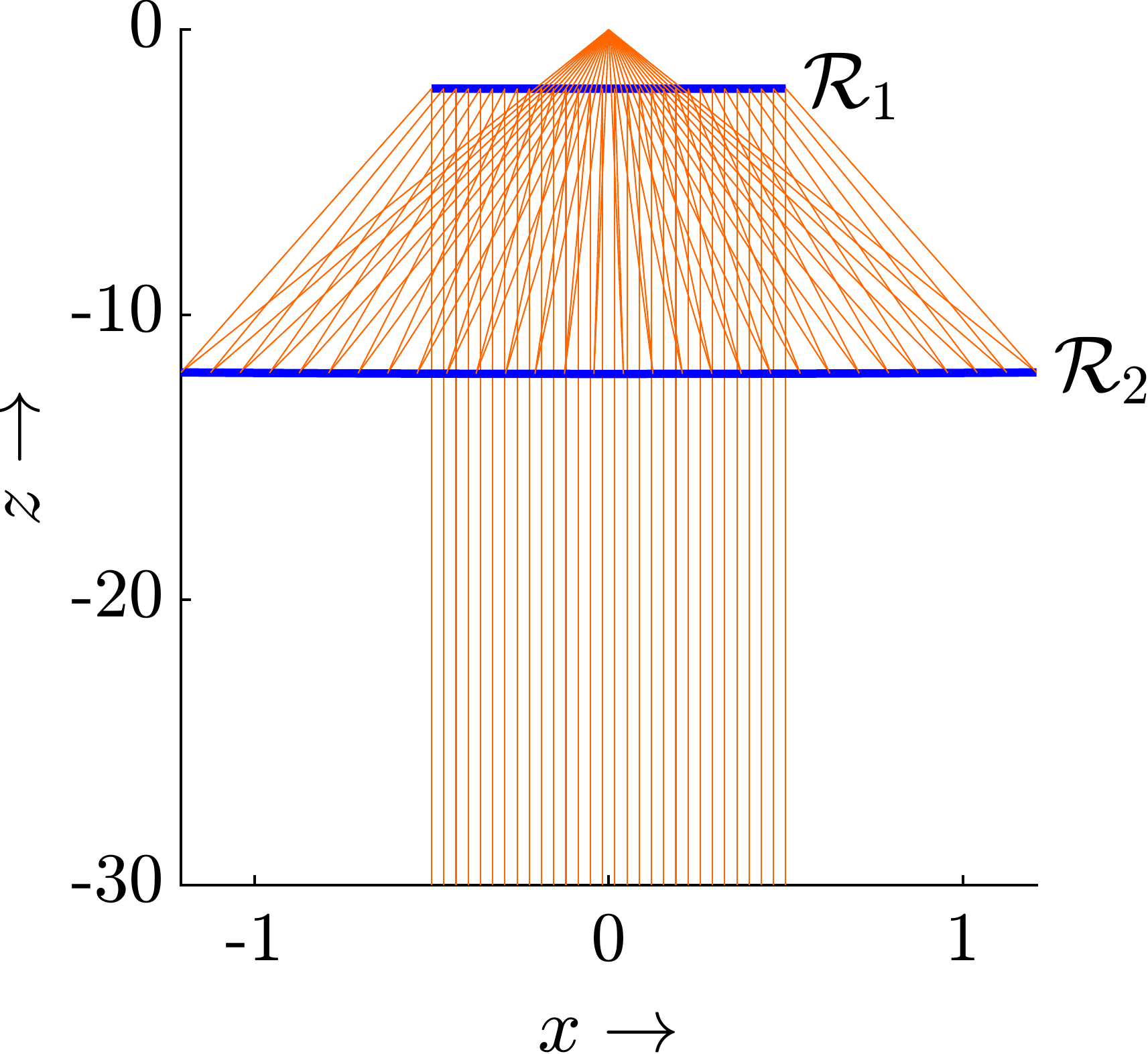}
        \caption{Layout of the optical system.}
        \label{fig: RT}
          \end{minipage}
          \begin{minipage}[b]{0.38\textwidth}
          \centering
              \includegraphics[trim={0cm 0cm 0cm 0cm},clip,width=0.8\textwidth]{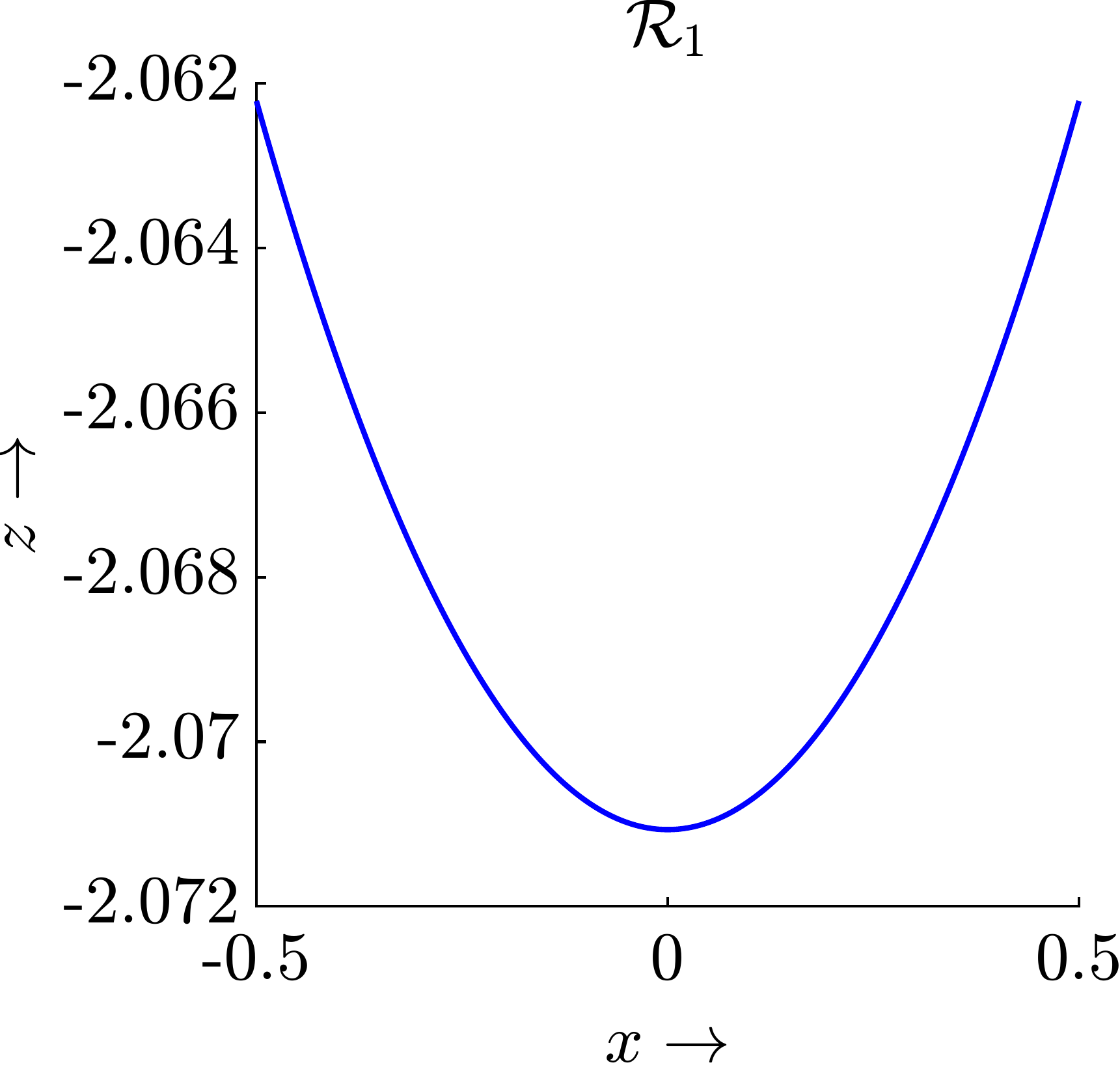}\\
              \vspace{1em}
              \includegraphics[trim={0cm 0cm 0cm 0cm},clip,width=0.8\textwidth]{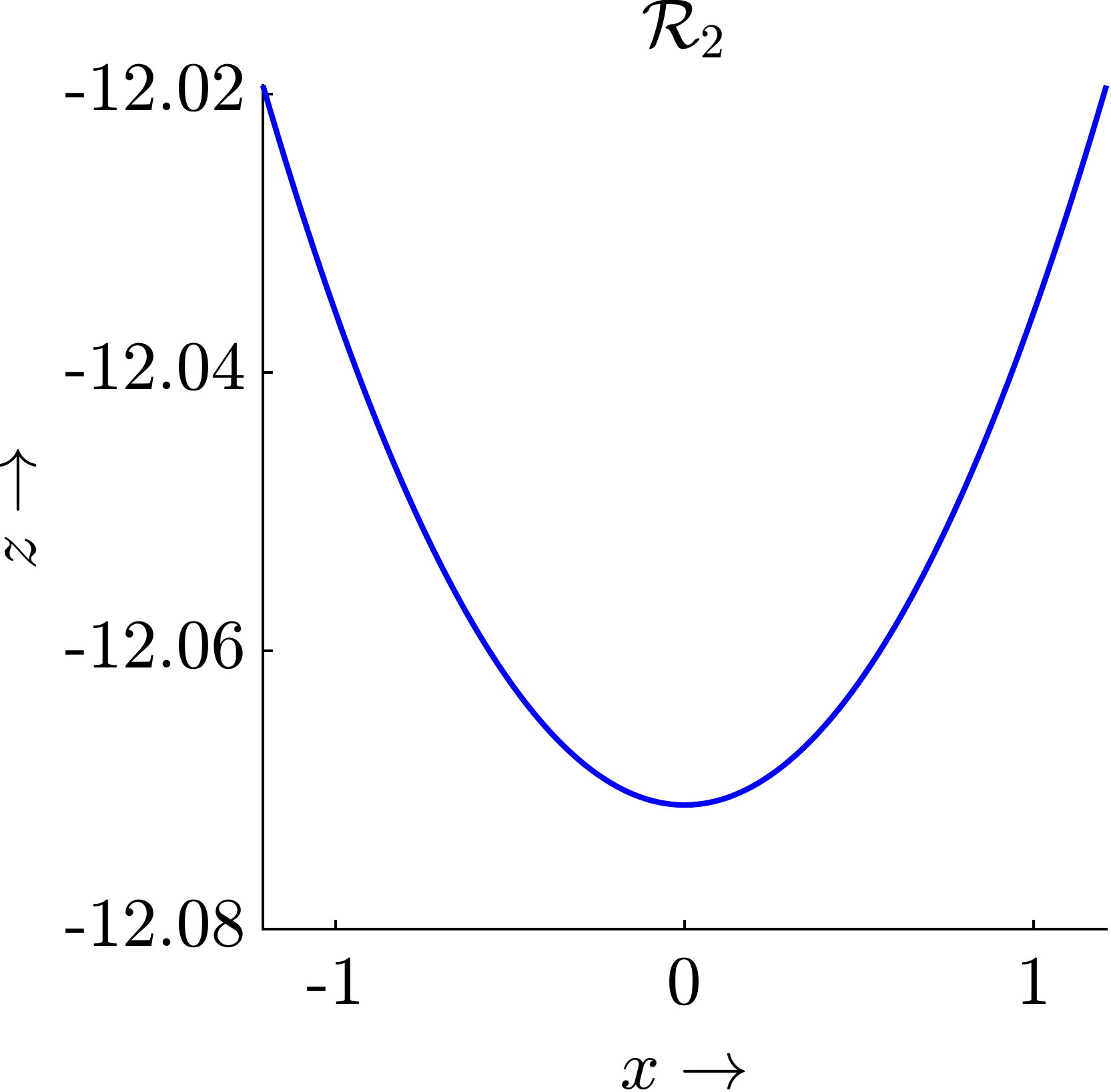}
         \caption{Shapes of the reflectors.}
          \label{fig: shapes}
           \end{minipage}
    \end{figure}

   We choose the parameters for the Schwarzschild design \cite{korsch} that are known for maximum correction of third-order aberrations. The shape of each reflector is given by
    \begin{equation}\label{eq: schwarz}
    z=\mathcal{R}_{i}(x)=\frac{x^2}{r_{i}+\sqrt{r_{i}-(1+C_{i})x^2}}, \quad i=1,2,
    \end{equation}
    where $C_i$ are the deformation constants, $r_i$ are the radii of curvature and subscript $i$ corresponds to the first and second reflectors. All parameters are defined by the relations given in Table~\ref{table: telescope parameters}, where $D_{\rm{v}}$ and $f_{\rm{s}}$ denote the distance between the vertices of both reflectors and 
    the focal length of the system, respectively. Note that, unlike the inverse design, both reflectors in the Schwarzschild design are functions of the source coordinates $x$. The layout of the optical system for both designs is shown in Fig.~\ref{fig: RT}. The shapes of both reflectors are given by Fig.~\ref{fig: shapes}.
    
    \begin{table}[htbp]
	\renewcommand{\arraystretch}{1.1}
	\setlength{\tabcolsep}{2.5cm}
	\caption{Parameters for a Schwarzschild telescope.}
	\normalsize
	\label{table: telescope parameters}
 \centering
	\begin{tabular}{c}
		\hline
		$\begin{array}{r@{\ } c@{\ } l}
			r_1&=&r_2=-2f_{\rm{s}}\sqrt{2}\\
			D_{\rm{v}}&=&-2f_{\rm{s}} \\
			C_1&=&(1+\sqrt{2})^2\\	
			C_2&=&(1+\sqrt{2})^{-2}
		\end{array}$
		\\
		\hline
	\end{tabular}
 \end{table}
 
    \begin{figure}[h]
		\centering
				\includegraphics[trim={0cm 0cm 0cm 0cm},clip, height=8cm]{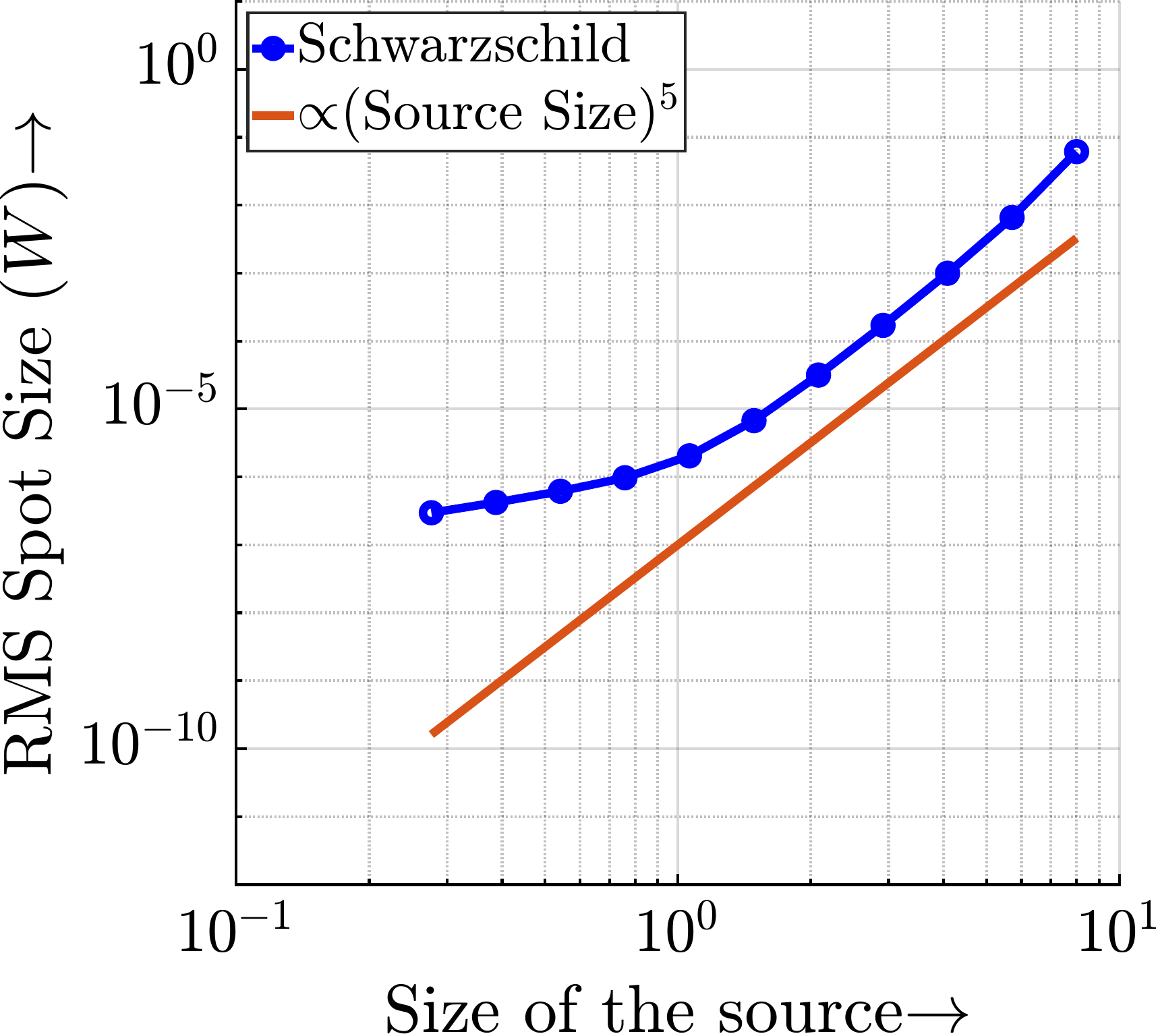}
    \caption{Schwarzschild design has fifth-order aberrations.}
        \label{fig: fifth order}
    \end{figure}
       
    We consider the source domain $Q_{\rm{s}}=[-0.5,0.5]$ and employ a raytracer in \textit{Matlab} based on the QI method described in Sect.~\ref{sec: raytracing}. We trace $10^4$ uniformly generated on-axis parallel rays with $\mathbf{\hat{s}}=(0,1)^T$ passing through the Schwarzschild telescope and find the target domain $P_{\rm{t}}=[- 0.09996, 0.09996]$. Both reflectors are approximated with $500$ points.
    
  In Fig.~\ref{fig: fifth order}, we compare the RMS spot size generated by $200$ on-axis rays from source domains of different lengths. We verify that the Schwarzschild telescope is limited by fifth-order aberrations.  Therefore, its target domain $P_{\rm{t}}$ is a suitable input parameter for the inverse freeform design. In Fig.~\ref{fig: fifth order}, numerical inaccuracies in the raytracer dominate for small source sizes.
     \begin{figure}[h]
    \begin{subfigure}{0.48\textwidth}    
        \centering
		\includegraphics[width=0.8\textwidth,height=5.5cm]{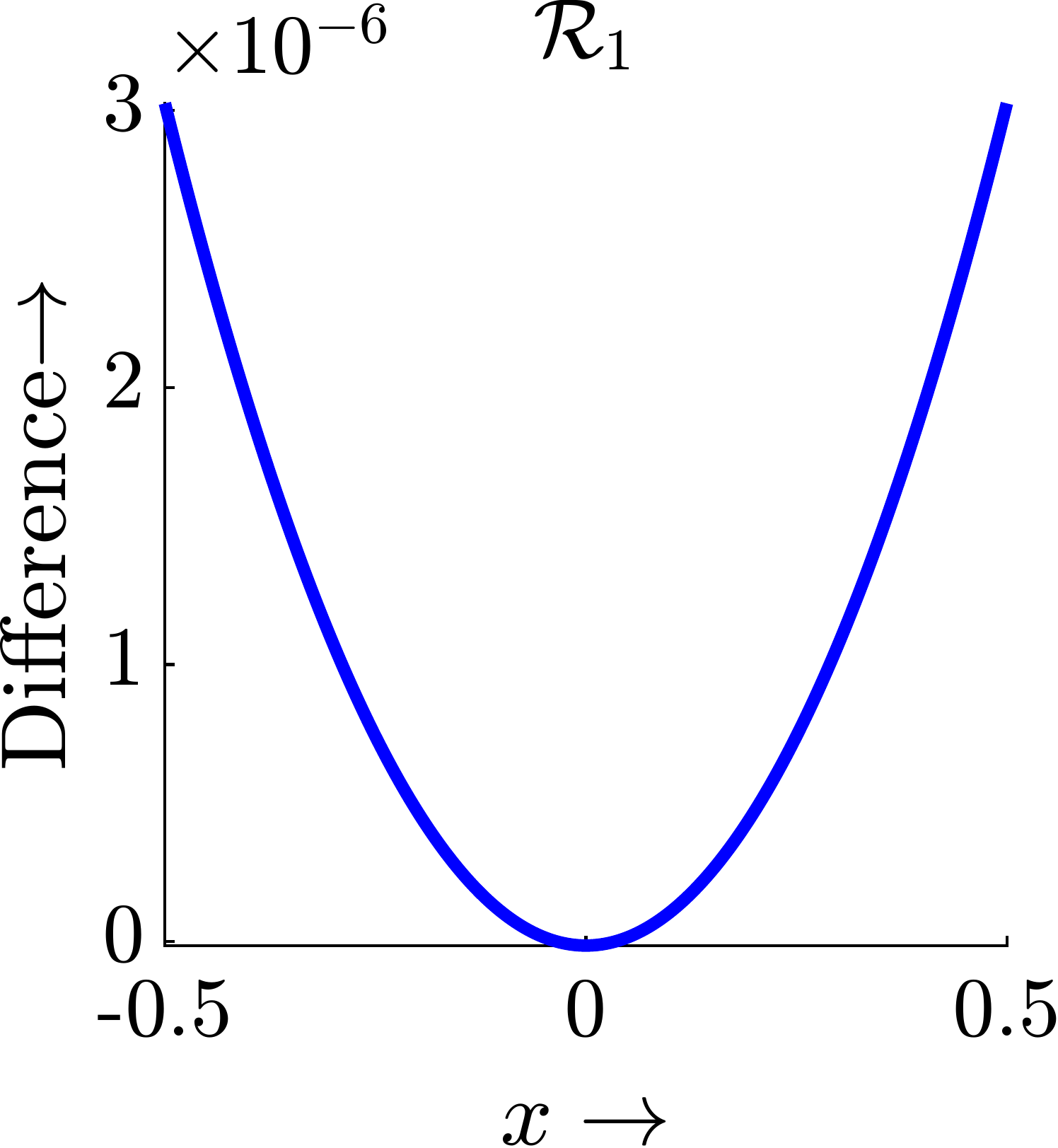}
        \subcaption{First reflector}
        \label{fig: R1 error}
    \end{subfigure}
 \hfill
    \begin{subfigure}{0.48\textwidth}    
        \centering
  	  \includegraphics[width=0.8\textwidth,height=5.5cm]{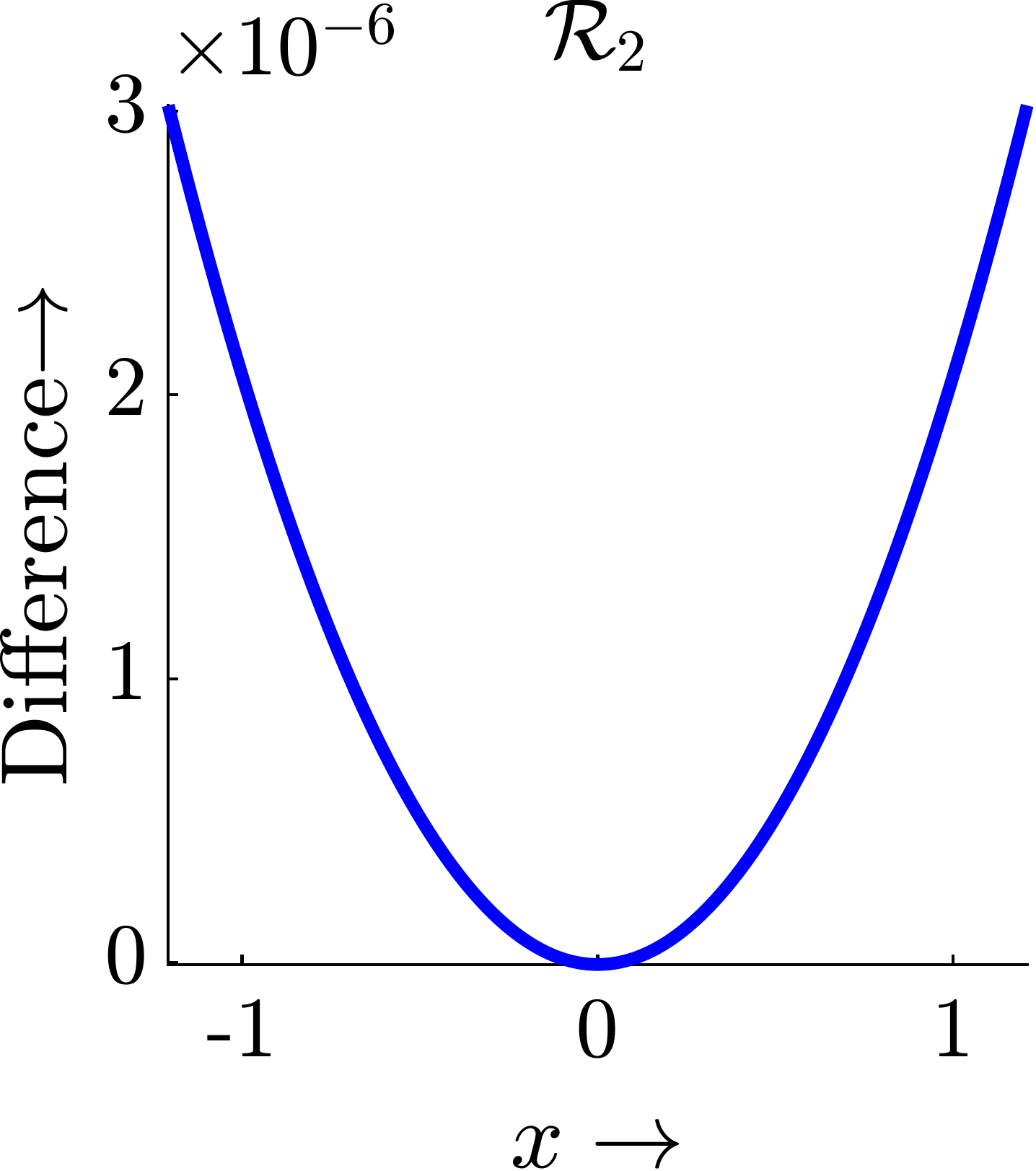}
        \subcaption{Second reflector}
        \label{fig: R2 error}
        \end{subfigure}
    \caption{Differences between the data points of the Schwarzschild and inverse freeform reflectors.}
    \label{fig: compare errors}
    \end{figure}
    \begin{figure}[htbp]
		\centering
    \begin{subfigure}{\linewidth}    
        \centering
				\includegraphics[trim={0cm 0cm 0cm 0cm},clip,height=5.5cm]{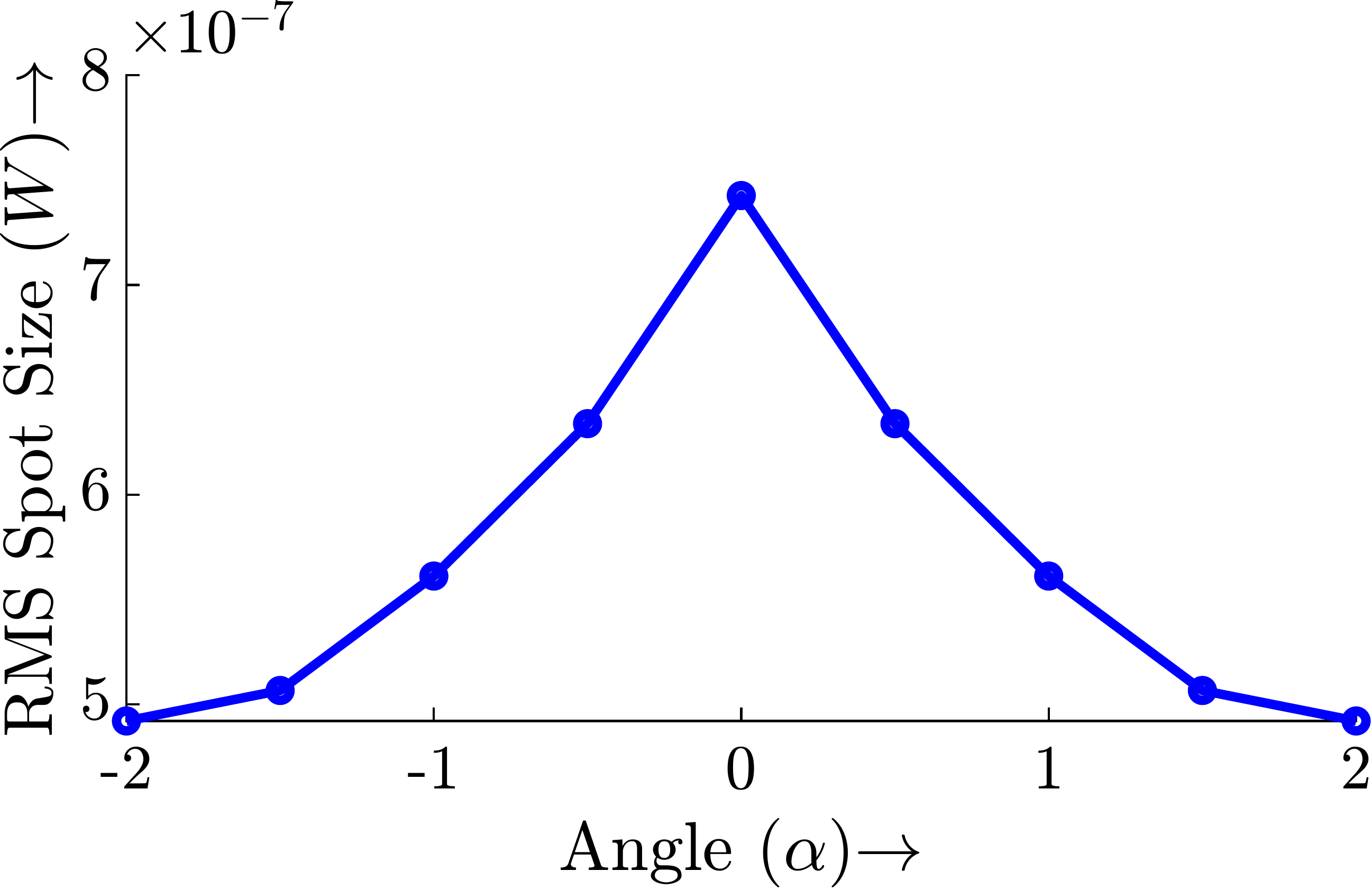}
    \caption{Schwarzschild design.}
        \label{fig:subfig-a}
    \end{subfigure}
    \vspace{1em}   
    \begin{subfigure}{\linewidth}    
        \centering
  	\includegraphics[trim={0cm 0cm 0cm 0cm},clip,height=5.5cm]{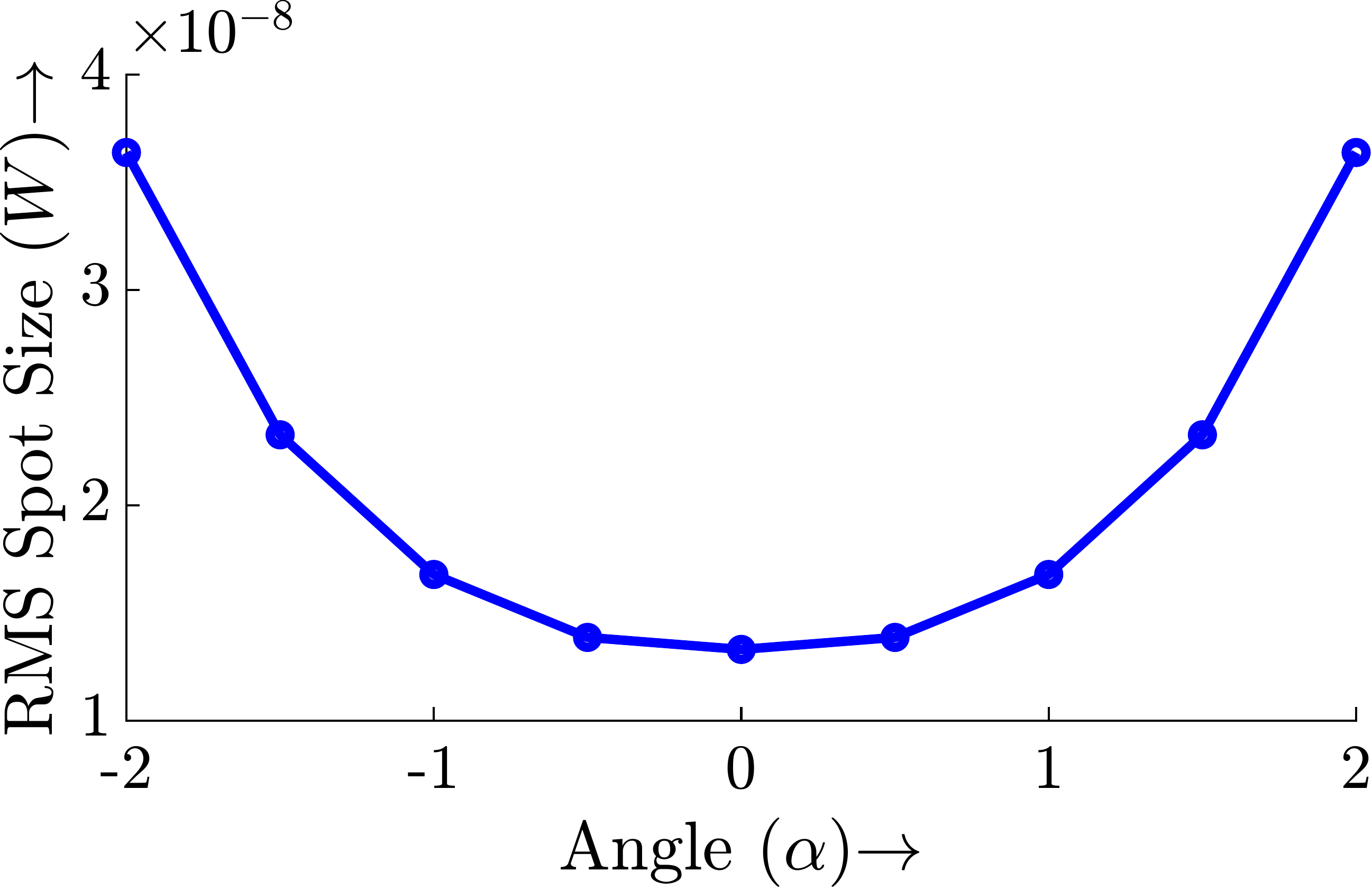}
   \caption{Inverse freeform design.}
        \label{fig:subfig-b}
    \end{subfigure}
    \caption{Comparison of RMS spot sizes $W$ for various angles $\alpha$.}
    \label{fig: compare spot size}
    \end{figure}

We will now compute the shapes of the inverse freeform imaging reflectors. Energy conservation on the above-mentioned source and target domains, $Q_{\rm{s}}$ and $P_{\rm{t}}$, respectively, is used to determine the ratio $k$ of the energy distributions. As elaborated in Sect.~\ref{sec: model}, we work with uniform energy distributions $f$ and $g$. We choose $f(q_{\rm{s}})=1$, and determine the total energy on the source of a unit length. Next, we find the value of $g(p_{\rm{t}})$ by energy conservation and the resulting value of $k$. With Eq.~(\ref{eq: phase-space linear map}), the phase-space optical map is obtained for a source grid with $500$ points. Subsequently, we compute the map $y=m(x).$ Then, as elaborated in Sect.~\ref{sec: model}, we compute $u_1(x)$ and both reflectors using Eqs.~(\ref{eq: necessary condition}) and (\ref{eq: reflector shapes}). The OPL and the distance of the vertex of the first reflector from the source are chosen as $V=50$ and $u_0=27.92,$ respectively.

The inverse freeform reflectors are compared with the Schwarzschild reflectors and the differences between the data points are shown in Fig.~\ref{fig: compare errors}. In Fig.~\ref{fig: compare spot size}, we present the RMS spot sizes when $200$ rays, each inclined at $\alpha\in\{-2^\circ,-1.5^\circ,\ldots,2^\circ\}$ to the optical axis pass through both systems. The RMS spot sizes for $\alpha=0^\circ$ for the Schwarzschild and inverse designs are $7.47e-07$ and $1.33e-08$, respectively. The Schwarzschild telescope is specifically designed to minimize off-axis aberrations, i.e., it is designed to achieve smaller spot sizes for $\alpha\neq0^\circ$. The RMS spot sizes corresponding to $\alpha=\pm2^\circ$ for the Schwarzschild and inverse designs are $4.92e-07$ and $3.63e-08$, respectively. We observe that although the inverse design is tailored to have an aberration-free image for on-axis rays, it still outperforms the classical Schwarzschild telescope for off-axis rays (see Table.~\ref{table: spot sizes}). It is interesting to note that small differences in the reflector data points lead to a distinct performance of both designs.
    \begin{table}[htbp]
	\renewcommand{\arraystretch}{1.1}
	\caption{RMS spot sizes for different angles.}
	\normalsize
	\label{table: spot sizes}
    \centering
	\begin{tabular}{|c|c|c|}
			\hline
				\textbf{Angle}& $0^{\circ}$& $\pm2^\circ$\\
				\hline
				\textbf{Classical Design} & $7.47e-07$ & $4.92e-07$ \\
				\textbf{Inverse Design} & $1.33e-08$ & $3.63e-08$ \\
				\hline
	\end{tabular}
\end{table}

   \section{Conclusion and future work}\label{sec: conclusion}
We proposed a design method to calculate freeform reflectors for a two-dimensional parallel-to-point double-reflector imaging system. We derived that the ratio of energy distributions at the source and target of an optical system must be constant for designing an aberration-free imaging system. This provided an approach to extend inverse methods for designing freeform nonimaging systems to imaging systems.

For input parameters like source and target domains, the inverse imaging design relies on the classical design that effectively minimizes aberrations. We verified the performance of our freeform double-reflector design by comparing the RMS spot sizes of on-axis and off-axis rays with those produced by the classical design. This required the implementation of a very accurate raytracer based on QI. QI can be easily extended to higher dimensions, which is necessary for designing three-dimensional (3D) systems in the future. We concluded that the inverse design is superior to the classical design for minimizing aberrations of off-axis rays. In commercial software, the design of imaging systems for diverse applications is based on optimization. The inverse freeform design may serve as a good starting point in optimization algorithms.

Inverse nonimaging methods establish a strong foundation for 3D freeform optical design \cite{lotte}. Future research may involve designing 3D inverse freeform imaging systems using the proposed method, enhancing its applicability.

\section*{Acknowledgments} 	The authors thank Teus Tukker (ASML) for testing and verifying the inverse design using CODE V. This research is supported by Topconsortium voor Kennis en Innovatie (TKI program ``Photolitho MCS" (TKI-HTSM 19.0162)).

\section*{References}
\bibliographystyle{iopart-num-modified}  
\bibliography{sample1}

\providecommand{\newblock}{}
\begin{thebibliography}{10}
\expandafter\ifx\csname url\endcsname\relax
  \def\url#1{{\tt #1}}\fi
\expandafter\ifx\csname urlprefix\endcsname\relax\def\urlprefix{URL }\fi
\providecommand{\eprint}[2][]{\url{#2}}

\bibitem{notes}
Wills S 2017 {\em Freeform Optics: Notes from the Revolution\/} {\em Opt.
  Photon. News\/} {\bf 28} 34--41

\bibitem{fabian}
Duerr F and Thienpont H 2021 {\em Freeform imaging systems: Fermat’s
  principle unlocks “first time right” design\/} {\em Light: Science \&
  Applications\/} {\bf 10} 95

\bibitem{janick}
Thompson K and Rolland J 2012 {\em Freeform Optical Surfaces: A Revolution in
  Imaging Optical Design\/} {\em Optics and Photonics News\/} {\bf 23} 30--35

\bibitem{SMS}
Mi{\~n}ano J~C, Ben{\'i}tez P, Lin W, Infante J, Mu{\~n}oz F and Santamar{\'i}a
  A 2009 {\em An application of the {SMS} method for imaging designs\/} {\em
  Opt. Express\/} {\bf 17} 24036--24044

\bibitem{korsch}
Korsch D 2012 {\em Reflective Optics\/} (Academic Press)

\bibitem{braat}
Braat J and Török P 2019 {\em Imaging Optics\/} (Cambridge University Press)

\bibitem{NAT}
Fuerschbach K, Rolland J~P and Thompson K~P 2014 Nodal aberration theory
  applied to freeform surfaces {\em Classical Optics 2014\/} (Optica Publishing
  Group)

\bibitem{opticaldesignsoftware}
Sahin F~E 2019 {\em Open-source optimization algorithms for optical design\/}
  {\em Optik\/} {\bf 178} 1016--1022 ISSN 0030-4026

\bibitem{evolutionary}
Nijkerk M~D, Gruber J~M and Boonacker B 2019 Freeform optics design tool for
  compact spectrometers {\em International Conference on Space Optics\/}

\bibitem{AIassistedopt}
Antonov K, Botari T, Tukker T, B{\"a}ck T, van Stein N and Kononova A~V 2023
  {New solutions to Cooke triplet problem via analysis of attraction basins}
  {\em Digital Optical Technologies 2023\/} vol 12624 (SPIE)

\bibitem{lotte}
Romijn L~B 2021 {\em Generated {J}acobian {E}quations in {F}reeform {O}ptical
  {D}esign: {M}athematical {T}heory and {N}umerics\/} Ph.D. thesis Eindhoven
  University of Technology

\bibitem{sanjana}
Verma S, Anthonissen M~J~H, ten Thije~Boonkkamp J~H~M and IJzerman W~L 2024
  {\em Design of two-dimensional reflective imaging systems: an approach based
  on inverse methods\/} {\em Journal of Mathematics in Industry\/} {\bf 14} 25

\bibitem{lagrangian}
Lakshminarayanan V, Ghatak A and Thyagarajan K 2002 {\em Lagrangian Optics\/}
  (Kluwer Academic Publishers)

\bibitem{symplecticdet1}
Rim D 2017 {\em An elementary proof that symplectic matrices have determinant
  one\/} {\em Adv. Dyn. Syst. Appl.\/} {\bf 12}

\bibitem{symplecticarea}
Abraham R, Marsden J~E and Ratiu T 2012 {\em Manifolds, {T}ensor analysis, and
  {A}pplications\/} vol~75 (Springer Science \& Business Media)

\bibitem{aberrationfree}
Wolf K~B 1988 {\em Symmetry-adapted classification of aberrations\/} {\em JOSA
  A\/} {\bf 5} 1226--1232

\bibitem{hecht}
Hecht E 2012 {\em Optics\/} (Pearson)

\bibitem{carmela}
Filosa C 2018 {\em Phase {S}pace {R}ay {T}racing for {I}llumination {O}ptics\/}
  Ph.D. thesis Eindhoven University of Technology

\bibitem{TBC}
Ries H and Rabl A 1994 {\em Edge-ray principle of nonimaging optics\/} {\em J.
  Opt. Soc. Am. A\/} {\bf 11} 2627--2632

\bibitem{yadav}
Yadav N~K 2018 {\em Monge-{A}mp\`{e}re {P}roblems with {N}on-{Q}uadratic {C}ost
  {F}unction: {A}pplication to {F}reeform {O}ptics\/} Ph.D. thesis Eindhoven
  University of Technology

\bibitem{evenorderaberrations}
Mori K, Hayasaki Y and Araki K 2021 {\em Fundamental-ray aberration analysis of
  off-axial optical systems: analytical formulae of first-order aberrations\/}
  {\em Appl. Opt.\/} {\bf 60} 9012--9028

\bibitem{QI}
Buhmann M and Jäger J 2022 {\em Quasi-Interpolation\/} Cambridge Monographs on
  Applied and Computational Mathematics (Cambridge University Press)

\bibitem{QIlychee}
Lyche T, Manni C and Speleers H 2017 B-splines and spline approximation
  \url{https://api.semanticscholar.org/CorpusID:195737484} (accessed on 13
  March 2025)

\end{thebibliography}

\end{document}